\documentclass[twocolumn,showpacs,pre, floatfix]{revtex4}

\begin{document}

\title{Nonlinear Elasticity of the Phase Field Crystal Model from the Renormalization Group}

\author{Pak Yuen Chan and Nigel Goldenfeld}
\affiliation{Department of Physics, University of Illinois at
Urbana-Champaign, Loomis Laboratory of Physics, 1110 West Green
Street, Urbana, Illinois, 61801-3080.}

\begin{abstract}
The rotationally-covariant renormalization group equations of motion
for the density wave amplitudes in the phase field crystal model are
shown to follow from a dynamical equation driven by an effective free
energy density that we derive.  We show that this free energy can be
written purely as a function of the strain tensor and thence derive the
corresponding equations governing the nonlinear elastic response.
\end{abstract}

% 05.70.Ln  Nonequilibrium and irreversible thermodynamics
% 81.15.Aa  Theory and models of film growth
% 81.16.Rf  Nanoscale pattern formation
% 05.10.Cc  Renormalization group methods
% 61.72.Cc  Kinetics of defect formation and annealing
% 62.20.Dc  Elasticity, elastic constants
% 81.40.Jj  Elasticity and anelasticity, stress-strain relations

 \maketitle

Multiple scale approaches to materials pattern formation are essential
to account for the variety of material structures that emerge on
different scales, and the dependence of materials properties on
structures at more than one scale\cite{Vvede04}.  However, a
coarse-grained description describing the pattern dynamics is not in
itself sufficient: for example, in fracture mechanics, one would like
to be able to predict the response of real, heterogeneous materials to
deformations, and to couple the resulting elasticity theory to
discrete, atomistic descriptions at small scales.  In short, the
challenge of materials modeling requires not just a coarse-grained
description of the underlying atomistic dynamics, but a
multi-resolution approach that can locally adapt to capture the
appropriate fast space and time scales.

Recently, we presented a proof-of-concept realization of such a
calculation\cite{Golden05,Badri06,athreya2007adaptive}, using a minimal
model of a crystalline material called the phase field crystal (PFC)
model\cite{Elder02,Elder04,provatas2007using}.The PFC model evolves
the local density field through a conservation law governed by a free
energy functional that penalizes departures from a perfectly periodic
ground state\cite{Brazovskii}.  The resulting density field retains the
crystallographic and elastic properties of the material, in contrast to
conventional phase field model descriptions of materials, and appears
to give a realistic account of a variety of phenomena, including
multicrystalline solidification\cite{Elder04},
elasticity\cite{Elder02,Stefa06,Elder04}, defect
dynamics\cite{Elder04,Berry06,Singe06}, epitaxial growth, as well as
crack and fracture dynamics\cite{Elder04}.  Our proof-of-concept
calculation consisted of two distinct steps: (i) a
renormalization-group calculation for the rotationally-covariant
equation of motion for the amplitudes of density waves corresponding to
the lattice periodicity, and (ii) the numerical solution of the
resulting equations using adaptive mesh refinement
techniques\cite{athreya2007adaptive}.  In two dimensions, this
calculation was three orders of magnitude faster than
straightforward integration of the PFC equations.

In this Rapid Communication we show that an additional level of
coarse-graining can be performed on the complex amplitude
representation, leading to a theory of nonlinear elasticity valid at
system-wide or macroscopic scales. Our construction is of interest
methodologically: instead of postulating a form for the strain energy
based on symmetry and phenomenology\cite{eringen1962ntc,atkin1980ite},
we derive it from a more microscopic model---in this case, the PFC
model.  Our work provides a theoretical connection to microscopic
structures and opens up the possibility of deriving other forms of
strain energy by similar approaches.  Finally, in varying our level of
description from the PFC model to the complex amplitude representation
to the strain energy formulation, we obtain the basis for a multiple
scale approach to materials properties, accessing the realm of
continuum mechanics from a density functional theory at atomic scales.
This approach could be applicable to interpreting recent experiments on
atomic force microscope nanoindentation of
graphene\cite{PhysRevB.64.235406,PhysRevB.69.115415}, complimenting
theoretical approaches using tight-binding
methods\cite{cadelano2009nonlinear}.

%The plan of this article is as follows: we first introduce the PFC
%model and its complex amplitude representation in the one-mode
%approximation, valid near threshold of the pattern-forming instability.
%We construct the corresponding free energy density in that
%representation, and show that the mode dynamics is not relaxational,
%because the overall dynamics of the density field is conservative, but
%is driven by an effective free energy whose form we calculate. The
%strain energy is then derived, by considering the Goldstone modes of
%the theory, and the resulting nonlinear elasticity free energy
%extracted.

\bigskip
\noindent
{\it The Model:-\/}
%\section{The Model}
The phase field crystal (PFC) model is defined by the free energy density\cite{Elder02,Elder04},
\begin{equation}
f = \frac{\rho}{2}(1+\nabla^2)^2\rho + \frac{r}{2}\rho^2 + \rho^4,
\label{eqn_f_rho}
\end{equation}
where $\rho(\vec{x},t)$ is the phase field, or the order parameter.
The dynamics is conservative and dissipative, given by
\begin{equation}
\frac{\partial \rho}{\partial t} = \nabla^2\left(\frac{\delta F}{\delta \rho}\right)\
\label{eqn_rho_dyn}
\end{equation}
where $F = \int d^2x f(x)$ is the total free energy of the system,
taken here to be two-dimensional. Thermal noise is ignored in our
discussion here, but could be included if desired.  Generally, it will
not be important at system scales. There are three phases in this
model, namely uniform, stripe and triangular.  The ground state in the
triangular phase can be written in the single-mode approximation by
\begin{equation}
\rho_{tri}(\vec{x}) = A\sum_{j=1}^3 \left(e^{i\vec{k}_j\cdot\vec{x}} + e^{-i\vec{k}_j\cdot\vec{x}}\right) + \rho_0,
\label{eqn_tri_gs}
\end{equation}
where $\rho_0$ is the average density, $A$ is the constant amplitude and
\begin{equation}
\vec{k}_1 = \hat{y},\quad
\vec{k}_2=\frac{\sqrt{3}}{2}\hat{x} - \frac{1}{2}\hat{y},\quad
\vec{k}_3=\frac{-\sqrt{3}}{2}\hat{x} - \frac{1}{2}\hat{y}
\label{eqn_k}
\end{equation}
are the three lattice vectors.  Goldenfeld {\it et al.}
showed\cite{Golden05, Badri06} that instead of using $\rho(\vec{x},t)$
as the dynamical variable, it is more efficient to generalize Eq.
(\ref{eqn_tri_gs}) and promote the constant amplitudes $A$ to $3$
slowly varying complex amplitudes $A_j(\vec{x},t)$ and treat the
$A_j(\vec{x},t)$ as dynamical variables.  Using
renormalization group techniques, they showed that the rotationally-covariant equation of motion for
$A_1(x,t)$ is given by\cite{Golden05,Badri06}
\begin{eqnarray}
\frac{\partial A_1}{\partial t} &=& (1-L_1)(\Gamma-L_1^2)A_1- 6\rho_0 A_2^*A_3^*\nonumber\\
&&-3A_1(|A_1|^2 + 2|A_2|^2 + 2|A_3|^2)+\cdots,
\label{eqn_A_1}
\end{eqnarray}
where $\Gamma\equiv -r -3\rho_0^2 $ and $L_1\equiv \nabla^2 +
2i\vec{k}_1\cdot\nabla$ is a rotationally covariant operator, and small
nonlinear gradient terms\cite{Badri06} are not written out explicitly.
The corresponding equations for $A_{2,3}(x,t)$ can be written down from the
appropriate permutations.

\bigskip
\noindent
{\it Derivation of the amplitude equations without using renormalization group:-\/}
%\section{Derivation of the Amplitude Equations without using Renormalization Group}
The first attempt to derive the free energy in this representation is
to note that Eq. (\ref{eqn_A_1}), ignoring the nonlinear gradient
terms, can be written as
\begin{equation}
\frac{\partial A_j}{\partial t} = -\frac{\delta F_{dyn}[A_j(x,t)]}{\delta A_j^*},
\label{eqn_A_dyn}
\end{equation}
where the free energy $F_{dyn}$ is given by the free energy density
\begin{eqnarray}
f_{dyn} &=&  -\sum_{j=1}^3 A_j^*(1-L_j)(\Gamma - L_j^2)A_j
+ 3\sum_{j,k=1}^3 |A_j|^2|A_k|^2 \nonumber\\
&&- \frac{3}{2}\sum_{j=1}^3 |A_j|^4 + 6\rho_0(A_1A_2A_3 + A_1^*A_2^*A_3^*),
\label{eqn_f_dyn}
\end{eqnarray}
The dynamics
given by Eq. (\ref{eqn_A_dyn}) is purely dissipative, as opposed to the
density-conserving dynamics in the original PFC equation, Eq.
(\ref{eqn_rho_dyn}).

%To resolve this conundrum, we note that Eq. (\ref{eqn_f_dyn}) is
%derived from the dynamical equations of $A_j$, which in turn, are
%derived from the PFC equation, Eq. (\ref{eqn_rho_dyn}), the
%density-conserving information in original PFC equation is propagated
%to the free energy, Eq. (\ref{eqn_f_dyn}).  However, density
%conservation should only be exhibited in the dynamical equation of
%motion and not be represented in the equilibrium free energy.  This can
%be seen in the original equations: density conservation appears only in
%the dynamical equation, Eq. (\ref{eqn_rho_dyn}), but not in the free
%energy, Eq. (\ref{eqn_f_rho}).

To resolve this conundrum, we note that mass
conservation should only be exhibited in the dynamical equation of
motion and not be represented in the equilibrium free energy.
Thus, the correct way to derive the free energy in the complex
amplitude representation is to derive it from the original PFC free
energy, Eq. (\ref{eqn_f_rho}).  The easiest way to do that is to
substitute the ansatz, Eq. (\ref{eqn_tri_gs}), into the free energy,
Eq. (\ref{eqn_f_rho}).  The first term of Eq. (\ref{eqn_f_rho}) can
then be computed by using the identity
\begin{equation}
(1+\nabla^2)\rho = \sum_{j=1}^3  \left(e^{i\vec{k_j}\cdot \vec{x}} L_jA_j + c.c.\right) + \rho_0,
\end{equation}
where $c.c.$ stands for complex conjugate.  By performing an integration by parts, we find
\begin{eqnarray}
\frac{\rho}{2}(1+\nabla^2)^2\rho
&=& \frac{1}{2}[(1+\nabla^2)\rho]^2\\
&=& \sum_{j=1}^3 A_j^*L_j^2A_j,
\label{eqn_tmp_step}
\end{eqnarray}
where in the last line constants and terms with the rapidly oscillating
factor $\exp{(i\vec{k}_j\cdot\vec{x})}$ are neglected.  We can neglect
the oscillating terms because the complex amplitudes, $A_j$, are slowly
varying on that scale, so the terms cancel themselves upon integration
over space.  Other terms in the free energy can be transformed in a
similar fashion. The resulting free energy is given by
\begin{eqnarray}
f_{amp}
&=& -\sum_{j=1}^3 A_j^* (\Gamma- L_j^2) A_j
+ 3\sum_{j,l=1}^3|A_j|^2|A_l|^2\nonumber\\
&&-\frac{3}{2}\sum_{j=1}^3 |A_j|^4
+ 6\rho_0(A_1A_2A_3 + A_1^*A_2^*A_3^*).
\label{eqn_f_amp}
\end{eqnarray}
Note that this free energy is different from Eq. (\ref{eqn_f_dyn}). The
$1-L_j$ operator in the first term in Eq. (\ref{eqn_f_dyn}) is absent
here.  This is to be expected because this operator arises from the
conservative Laplacian in the dynamical equation, and according to our
discussion above, it should not appear in the free energy.  The
transformation of the dynamical equation, Eq. (\ref{eqn_rho_dyn}), into
the complex amplitude representation can be performed by observing that
for any function, $f(x)$, the identity,
\begin{equation}
\int_{-\infty}^{\infty} d^2 x [(L_j-1)f(x)]e^{i\vec{k}_j\cdot\vec{x}} \equiv 0,
\label{eqn_conserved_int_A}
\end{equation}
holds, a counterpart to the identity,
\begin{equation}
\int_{-\infty}^{\infty} d^2 x \nabla^2 g(x) = 0,
\label{eqn_conserved_int_flux}
\end{equation}
for any function $g(x)$.  In fact, if we define
$g(x)=f(x)e^{i\vec{k}_j\cdot\vec{x}}$, Eq.
(\ref{eqn_conserved_int_flux}) implies Eq. (\ref{eqn_conserved_int_A}).
This shows that when we make the change of variables from the density,
$\rho$, to the complex amplitudes, $A_j$, the Laplacian in the
conservative dynamical equation has also to be transformed to $L_j-1$.
We thus arrive at the equation of motion
\begin{equation}
\frac{dA_j}{dt} = (L_j - 1)\frac{\delta F_{amp}}{\delta A_j^*},
\end{equation}
which, when written out explicitly, is,
\begin{eqnarray}
\frac{d A_1}{dt} &=& (1-L_1)\big[(\Gamma -L_1^2)A_1 - 6\rho_0 A_2^*A_3^*,\nonumber\\
&&- 3A_1(|A_1|^2 + 2|A_2|^2 + 2|A_3|^2 )\big]
\label{eqn_amp1_conserved}
\end{eqnarray}
with appropriate permutations for $A_{2,3}(x,t)$.  By construction,
these equations conserve the density of the system.

Eq. (\ref{eqn_amp1_conserved}) is exactly the same as Eq.
(\ref{eqn_A_1}) with all the nonlinear gradient terms included. This
derivation shows that the inclusion of the nonlinear gradient terms is
crucial for density conservation, and that all those terms can actually
be written in the condensed form of Eq. (\ref{eqn_amp1_conserved}).
Note, however, that our derivation does not explicitly use a
renormalization group argument, but follows from the integral identity,
Eq. (\ref{eqn_conserved_int_A}).  The connection with the renormalization
group approach arises from the coarse-graining assumption
after Eq. (\ref{eqn_tmp_step}), where we asserted that the
rapid oscillation averages to zero.

%Eq. (\ref{eqn_amp1_conserved}) is different from Eq. (\ref{eqn_A_1}).  Upon comparison, we find that there are many extra nonlinear gradient terms in Eq. (\ref{eqn_amp1_conserved}), because the operator, $1-L_1$, is acting on all the terms on the right, instead of just the first term.  It is important to observe that these extra nonlinear gradient terms are exactly the same extra terms we would get if we did the correct operator ordering in the derivation of the complex amplitude equations\cite{Badri06}.  Recall that Eq. (\ref{eqn_A_1}) is derived by a \lq quick and dirty' method, in which the complex amplitudes are renormalized \textit{after} the conservative Laplacian operator acts on them.  Athreya {\it et al.} has already showed that this method misses all the nonlinear gradient terms\cite{Badri06} that would be generated if other systematic asymptotic methods, such as the Renormalization Group and multiple scale analysis, were used.  Our analysis, however, complements their argument and show that these nonlinear gradient terms are indeed non-negligible.  In fact, the complex amplitude equations without these terms violet density conservation.  Our analysis also provide another quick way to write down the full amplitude equations---by expanding the free energy in the amplitudes and replacing the conservative Laplacian with the $L_j - 1$ operator.

%With this correct free energy in the complex amplitude representation,
%we can now proceed and derive the nonlinear elastic properties of the
%model.

\bigskip
\noindent
{\it Nonlinear Elasticity:-\/}
%\section{Nonlinear Elasticity}
The goal of this section is to derive the free energy as a function of
the strain tensor, when the PFC crystal is deformed under a general
deformation
\begin{equation}
x_m' = F_{mn}x_n,
\label{eqn_xFx}
\end{equation}
where $F_{mn}$ is the deformation gradient.  Einstein's summation
convention is used throughout, except for the index $j$ in $\vec{k}_j$,
$A_j$ and $L_j$. In general, the deformation gradient can be written
as\cite{atkin1980ite}
\begin{equation}
F_{mn}=R_{mp}U_{pn}
\end{equation}
where $R_{mp}$ is a pure rotation matrix and $U_{pn}$ is a
positive-definite, pure deformation matrix.  Since our system is
rotationally covariant, we expect that the free energy should only
depend on the function $U^TU$, where $U^T$ is the transpose of the
matrix $U$.

Under the deformation, Eq. (\ref{eqn_xFx}), the complex amplitudes transform as
\begin{equation}
A_j\rightarrow A_j'=A e^{ik_{jm}D_{mn}x_n},
\label{eqn_A_deform}
\end{equation}
where we defined $D_{mn}\equiv R_{mk}U_{kn}-\delta_{mn}$ and assumed
that $|A_j|=A$ for all $j$, where $A$ is a constant.  $k_{jm}$ is the
$m$-th component of the vector $\vec{k}_j$.  Because the deformation
gradient only enter the complex amplitude through its phase, describing
local density deformations, the only relevant terms in the free energy
are the gradient terms given by
\begin{equation}
E \equiv \sum_{j=1}^3 A_j^*L_j^2A_j.
\label{eqn_E2}
\end{equation}
Other terms in the free energy, Eq. (\ref{eqn_f_amp}), only contribute
when we minimize the free energy with respect to $A$ at the end of the
calculation.  By using Eq. (\ref{eqn_A_deform}) and differentiating, we
obtain
\begin{equation}
L_jA_j = (-k_{jm} k_{jn} R_{mp} R_{na} U_{pq} U_{aq} + 1)A_j.
\end{equation}
Apply $L_j$ again and substitute the result into Eq. (\ref{eqn_E2}) to obtain
\begin{equation}
E = A^2(E_1-2E_2+3),
\label{eqn_E}
\end{equation}
where
\begin{equation}
E_1 = \left(\sum_{j=1}^3 k_{jm} k_{jn}k_{ju} k_{jv}\right)F_{mq}F_{nq}F_{uw}F_{vw},
\label{eqn_E1}
\end{equation}
and,
\begin{equation}
E_2 = \left(\sum_{j=1}^3 k_{jm} k_{jn}\right)F_{mq}F_{nq}.
\end{equation}
The rest of the derivation concerns the evaluation of $E_1$ and $E_2$.
We first evaluate $E_2$.  By using the definition of $\vec{k}_j$ from
Eq. (\ref{eqn_k}), we obtain
\begin{equation}
k_{1m}k_{1n} = \delta_{my}\delta_{ny} = \delta_{mn}\delta_{my},
\end{equation}
\begin{equation}
k_{2m}k_{2n} =
\frac{3}{4}\delta_{mx}\delta_{nx}+\frac{1}{4}\delta_{my}-\frac{\sqrt{3}}{4}(\delta_{mx}\delta_{ny}+\delta_{nx}+\delta_{my}),
\end{equation}
and,
\begin{equation}
k_{3m}k_{3n} =
\frac{3}{4}\delta_{mx}\delta_{nx}+\frac{1}{4}\delta_{my}+\frac{\sqrt{3}}{4}(\delta_{mx}\delta_{ny}+\delta_{nx}+\delta_{my}).
\end{equation}
Combining these three equations we have
\begin{equation}
\sum_{j=1}^3 k_{jm} k_{jn} =
\frac{3}{2}\delta_{mn}(\delta_{mx}+\delta_{my}) =
\frac{3}{2}\delta_{mn}.
\end{equation}
Thus, $E_2$ is given by
\begin{equation}
E_2= \frac{3}{2}F_{pq}F_{pq}=\frac{3}{2} U_{pq}U_{pq} = \frac{3}{2} \texttt{Tr}[U^T U],
\end{equation}
where we used the property of the rotation matrix, that
$R_{im}R_{jm}=R_{mi}R_{mj}=\delta_{ij}$, $\texttt{Tr}[A]$ and $A^T$ are
the trace and transpose of the matrix $A$ respectively.  The evaluation
of $E_1$ is more involved.  We note that
\begin{equation}
k_{1m} k_{1n}k_{1u} k_{1v} = \delta_{mn}\delta_{nu}\delta_{uv}\delta_{my},
\end{equation}
and observe that the term
\begin{equation}
k_{2m}k_{2n}k_{2u} k_{2v} +k_{3m} k_{3n}k_{3u} k_{3v}
\end{equation}
is equal to the term
\begin{equation}
2\times[k_{2m}k_{2n}k_{2u} k_{2v} + \texttt{ terms with positive coefficients}].
\end{equation}
By exploring this relation and using the definition of $\vec{k}_j$, Eq. (\ref{eqn_k}), we obtain
\begin{eqnarray}
\sum_{j=1}^3 k_{jm} k_{jn}k_{ju} k_{jv}
&=& \frac{3}{8}(\Delta_{xxyy}+\Delta_{yyxx}+\Delta_{xyxy})\nonumber\\
&&+
\frac{3}{8}(\Delta_{xyyx}+\Delta_{yxxy}+\Delta_{yxyx})\nonumber\\
&&+
\frac{9}{8}(\Delta_{xxxx}+\Delta_{yyyy}),
\label{eqn_k4}
\end{eqnarray}
where we defined $\Delta_{abcd}\equiv\delta_{ma}\delta_{nb}\delta_{uc}\delta_{vd}$ for convenience.  By this, we have
\begin{eqnarray}
&&\left(\sum_{j=1}^3 k_{jm} k_{jn}k_{ju} k_{jv}\right)
R_{mp} R_{na}R_{us} R_{vt}\nonumber\\
&=&
\frac{3}{8}[R_{xp}R_{xa}R_{ys}R_{yt} + R_{yp}R_{ya}R_{xs}R_{xt}]\nonumber\\
&&+
\frac{3}{8}[R_{xp}R_{ya}R_{xs}R_{yt} + R_{yp}R_{xa}R_{ys}R_{xt}]\nonumber\\
&&+
\frac{3}{8}[R_{xp}R_{ya}R_{ys}R_{xt} + R_{yp}R_{xa}R_{xs}R_{yt}]\nonumber\\
&&+ \frac{9}{8}[R_{xp}R_{xa}R_{xs}R_{xt} + R_{yp}R_{ya}R_{ys}R_{yt}]
\label{eqn_k4_1}
\end{eqnarray}
To evaluate this expression, we note that we can combine terms
judiciously.  For example, by using the property of the rotation
matrix, we obtain
\begin{equation}
R_{xp}R_{xa}R_{ys}R_{yt}  + R_{xp}R_{xa}R_{xs}R_{xt} = R_{xp}R_{xa}\delta_{st},
\label{eqn_int2}
\end{equation}
and,
\begin{equation}
R_{yp}R_{ya}R_{xs}R_{xt} + R_{yp}R_{ya}R_{ys}R_{yt} =
R_{yp}R_{ya}\delta_{st}.
\label{eqn_int3}
\end{equation}
The sum of Eq. (\ref{eqn_int2}) and (\ref{eqn_int3}) then give
\begin{eqnarray}
&&R_{xp}R_{xa}R_{ys}R_{yt}  + R_{xp}R_{xa}R_{xs}R_{xt} \nonumber\\
&&+ R_{yp}R_{ya}R_{xs}R_{xt} + R_{yp}R_{ya}R_{ys}R_{yt}
= \delta_{st}\delta_{ap}
\end{eqnarray}
Repeating this for all the terms in Eq. (\ref{eqn_k4_1}), we obtain
\begin{eqnarray}
&&\left(\sum_{j=1}^3 k_{jm} k_{jn}k_{ju} k_{jv}\right)
R_{mp} R_{na}R_{us} R_{vt}\nonumber\\
&=&
\frac{3}{8} [\delta_{ap}\delta_{st} + \delta_{sp}\delta_{at} + \delta_{as}\delta_{pt}].
\end{eqnarray}
Substituting into Eq. (\ref{eqn_E1}), and then into Eq. (\ref{eqn_E}), we obtain
%\begin{equation}
$E = 3A^2\Delta$,
%\end{equation}
where
\begin{equation}
\Delta = \frac{1}{8}\big\{[\texttt{Tr}(U^TU)]^2 +2 \texttt{Tr}(U^TUU^TU)\big\} - \texttt{Tr}(U^T U) + 1.
\end{equation}
By using the relation
$
U^T_{mp}U_{pn} = \delta_{mn} + 2 u_{mn},
$
where $u_{ij}\equiv (\partial _i u_j + \partial _j u_i + (\partial _k
u_i)(\partial _k u_j))/2$ is the nonlinear strain tensor, we obtain
\begin{equation}
\Delta = \left(\frac{3}{2}u_{xx}^2 + \frac{3}{2}u_{yy}^2 + u_{xy}^2 + u_{yx}^2 + u_{xx}u_{yy}\right).
\end{equation}

Finally, we substitute back into the free energy, Eq. (\ref{eqn_f_amp}),
and minimize the whole expression with respect to $A$ to obtain
\begin{equation}
A(\Delta) = \frac{1}{5}\left(-\rho_0 \pm \frac{1}{3}\sqrt{9\rho_0^2 + 15(\Gamma - \Delta)}\right),
\end{equation}
which gives the free energy density as
\begin{equation}
f(\Delta) = \frac{45}{2}A^4(\Delta) + 12\rho_0A^3(\Delta) - 3(\Gamma-\Delta)A^2(\Delta).
\label{final_result}
\end{equation}
This formula completely defines the elastic properties of the PFC
model, and provides a starting point for conventional continuum
mechanical applications of nonlinear elasticity theory.

%The connection
%to linear elasticity theory is made by expanding the free energy around
%the ground state, giving
%\begin{equation}
%f-f_0
%= 3 A_0^2 \Delta + \frac{3\rho_0A_0 - \Gamma}{6\rho_0^2 + 10\Gamma}\Delta^2 + O(\Delta^3),
%\label{eqn_nonlinear_fA}
%\end{equation}
%where $f_0$ is the ground state energy and the first order term of this
%equation is the usual elastic energy of an isotropic medium.

%We conclude with some brief remarks about the possible ways our results
%could be used for describing real materials exhibiting
%polycrystallinity and defects.

Nonlinear elastic theories are coarse-grained models up to the level of
the continuum, and so do not explicitly include defect structures; our
results are most useful for understanding large deformation behaviour,
twinning and phase transitions in ordered materials.  However, for
plastic deformations, dislocations need to be included, usually by
postulating a free energy with contributions from nonlinear elasticity
(such as Eq. (\ref{final_result})), vacancies and
dislocations\cite{onuki2003plastic,minami2007nonlinear}.  While this
can yield useful insights, the most suitable level of description for
probing the multi-scale phenomena accompanying plastic flow remains the
PFC equations taking into account vacancies\cite{chan2009molecular} or
their rotationally-covariant renormalized counterparts (if only
dislocations are present).  We will report on this approach in a future
publication\cite{CHAN2010}.

%To summarize, we have shown how the phase variation in the
%renormalization group equation of the PFC model yield a complete
%derivation of nonlinear elasticity, one whose parameters can be related
%to microscopic parameters present in the original PFC equation.

\bibliographystyle{apsrev}

\bibliography{pfc_elasticity_bib}

\end{document}